\documentclass{IEEEoj}
\usepackage{cite}
\usepackage{amsmath,amssymb,amsfonts}
\usepackage{algorithmic}
\usepackage{graphicx}
\usepackage{textcomp}
\usepackage{amsmath}
\usepackage{amssymb}
\usepackage{graphicx}
\usepackage{cite}
\usepackage{amsthm}
\usepackage{cleveref}
\usepackage{array}
\usepackage{multirow}
\usepackage{makecell}
\usepackage{cite}
\usepackage{siunitx}
\usepackage{float}

\usepackage{verbatim}
\begin{document}
\receiveddate{XX Month, XXXX}
\reviseddate{XX Month, XXXX}
\accepteddate{XX Month, XXXX}
\publisheddate{XX Month, XXXX}
\currentdate{XX Month, XXXX}
\doiinfo{OJAP.2020.1234567}
\def\OJlogo{\vspace{-10pt}\includegraphics[height=20pt]{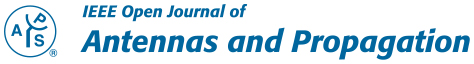}}
\title{Path Loss, Angular Spread and Channel Sparsity Modeling for Indoor and Outdoor Environments at the sub-THz Band}
\author{\uppercase{Dimitrios~G.~Selimis}\authorrefmark{3}, \IEEEmembership{Member, IEEE},
\uppercase{Mar Francis De Guzman}\authorrefmark{2},~\IEEEmembership{Student Member, IEEE}, \uppercase{Fotis~I.~Lazarakis}\authorrefmark{1},
\uppercase{Kyriakos~N.~Manganaris}\authorrefmark{1}, \uppercase{Kostas~P.~Peppas}\authorrefmark{3}~\IEEEmembership{Senior~Member,~IEEE}  and \uppercase{Katsuyuki Haneda}\authorrefmark{2}~\IEEEmembership{Member, IEEE}}
\affil{Institute of Informatics and Telecommunications, National Center for Scientific Research ``Demokritos,'' 15310 Athens, Greece  (e-mail:\{flaz,\{dselimis\}@iit.demokritos.gr}
\affil{Aalto University-School of Electrical Engineering, Dept. of Electronics and Nanoengineering, Espoo, Finland,
email: \{francis.deguzman,katsuyuki.haneda\}@aalto.fi}
\affil{Department of Informatics and Telecommunications, University of Peloponnese, 22131 Tripoli, Greece (e-mail: \{dselimis, peppas\}@uop.gr),}


\corresp{Corresponding author: Kostas~P.~Peppas (e-mail: peppas@uop.gr)}
\markboth{Path Loss, Angular Spread and
Channel Sparsity Modeling for Indoor
and Outdoor Environments at the
sub-THz Band}{D. Selimis \textit{et al.}}
\begin{abstract}
In this paper, we present new measurement results to model large-scale path loss, angular spread and channel sparsity at the sub-THz (141-145 GHz) band, for both indoor and outdoor scenarios. Extensive measurement campaigns have been carried out, taking into account both line-of-sight (LoS) and non line-of-sight (NLoS) propagation. For all considered propagation scenarios, omni-directional and directional path loss models have been developed, based on the so-called close-in (CI) free-space reference distance model. Moreover, path loss modeling has been applied for the $2^{\text{nd}}$ and $3^{\text{rd}}$ strongest multipath components (MPCs), based on which path loss exponent and large-scale shadow fading estimates have been derived. A power angular spread analysis is further presented, using up to the $3^{\text{rd}}$ strongest MPC. Finally, results on the sparsity of the wireless channel have also been presented by employing the so-called Gini index.  
\end{abstract}



\maketitle

\section{Introduction}\label{Sec:Intro}
\IEEEPARstart{S}ub-THz communications systems, operating in the frequency range between 100 GHz and the beginning of THz band, have recently attracted significant attention within both the wireless research community as well as for commercial purposes. The ever increasing demand for extreme high data rates has already pushed the current sub-6 GHz telecommunication systems to their limits. Sub-THz communications are therefore envisaged to fulfill the requirements for high data-rate communications and successfully complement telecommunication systems operating at lower frequency bands \cite{RappaportArticle,Katsu}. 

Accurate modeling of channel and propagation characteristics is crucial for the performance analysis and design of telecommunication systems operating at such bands. However, to the best of our knowledge, very few studies on channel and propagation characteristics for frequencies above 100 GHz are available in the open technical literature.  
In particular, a number of important wireless propagation parameters, including path loss, gas attenuation and diffraction, require in-depth research. 
Among others, large-scale path loss analysis and modelling for indoor as well as outdoor
environments are of significant importance. Representative examples can be found in \cite{C:XingRapaport,Xing,Joonas,Cheng,HeJia, Ju2,JuRappaport,Nguyen,Pometcu2,MacCartney}.

Most of the existing past research works deal with path-loss modeling in indoor environments. For example, in \cite{Xing} frequencies up to 140 GHz are evaluated for an indoor scenario. A similar approach has been adopted in \cite{HeJia}, assuming a meeting and an office room, for operating frequencies of 140 GHz and 220 GHz, respectively, and directional antennas. Lastly, in \cite{Joonas},\cite{Cheng} signal propagation at 10, 140 and 300 GHz has been addressed.
\begin{figure*}[ht]
\centering\vspace{0.5cm}
\includegraphics[keepaspectratio,width=\linewidth]{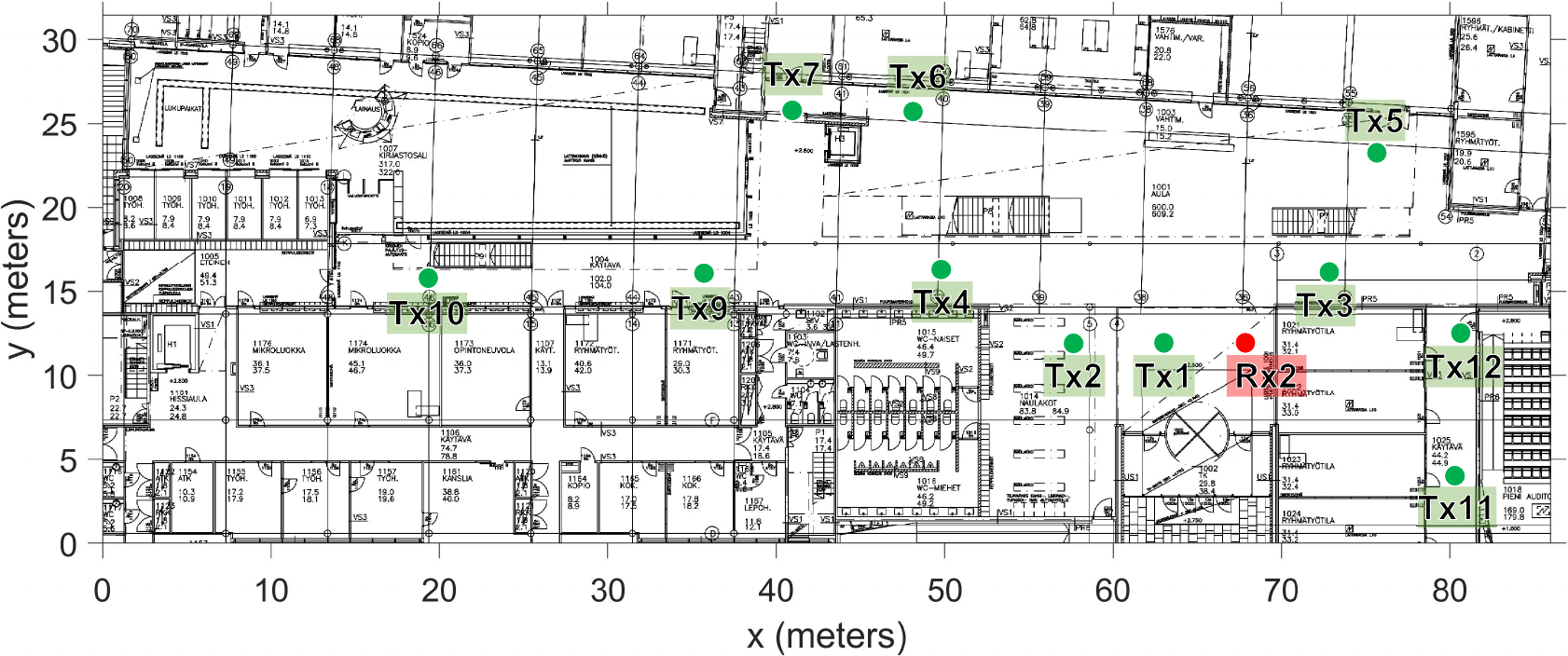}
\caption{Antenna locations of the first measurement campaign in the entrance hall}\vspace{-0.4cm} \label{fig:TUAS_map1}
\end{figure*}
Regarding the outdoor propagation cases, there are relatively few studies available in the literature. Specifically, in \cite{C:XingRapaport,Rappaport2021,JuConference,ChenYi,DeBeelde} measurement campaigns in urban, suburban micro-cells with path loss modeling at 142 GHz are described. Extensive works consisting of  both indoor and outdoor sets, estimating large scale fading are not many for the present time. As an example, authors in \cite{Khatun} calculate path loss exponent at $73$ GHz, for indoor and outdoor Airport environments, using only the  LoS links.

On the other hand, power angular spread can provide useful information, which, in turn, can be exploited to predict the angle of arrival (AoA) of each multipath component (MPC). In several cases of interest, power angular spread can be employed in an efficient manner to estimate the receiver-transmitter antenna distance. 
For example, in \cite{TPeng}, power angular profile data has been utilized to predict the arrival angles in a lecture room scenario for frequencies up to 300 GHz. In ~\cite{J:Zhang}, power angular profile has been used to provide information about the correlation of transmit-receive antenna distance, depending on the surrounding environment, as well as the propagation conditions.  

Finally, channel sparsity has been recently proposed as an efficient metric to provide further insights on the nature of wireless channels for future intelligent communication systems. 
Intuitively, the sparsity of a channel serves as a representation of the concentration of the power of a signal. 
In general, a wireless channel is assumed to be sparse when the number of observed MPCs is small \cite{C:Rusu, HurleyGini}. In such a channel, few signal MPCs are of high energy i.e 
channel is dominated by a small number of significant
paths. The energies of the MPCs are 
distributed into small regions in angle and delay and the wireless channel is sparse \cite{ZhangGini}.

In order to quantify the sparsity of a wireless channel, a suitable sparsity measure is required. Such metrics have been inspired by economic sciences when analysing the inequities of wealth \cite{DaltonGini}. Although several measures have been proposed in the past for estimating sparsity, the so called Gini index\cite{Gini} has attracted significant attention within the research community. Recently, this metric has been successfully employed to predict channel sparsity. Representative examples can be found in \cite{ZhangGini,HurleyGini,ZonoobiGini,T:Hurley2} and references therein.
Motivated by the above, the main contributions of this work can be summarized as follows. 
\begin{itemize}
    \item We investigate path loss modeling at 140 GHz based on extensive channel measurement campaigns conducted in indoor and outdoor environments, including both LoS and NLoS links. 
Measurements have been conducted by employing frequency-domain channel
sounding method via a vector network analyzer (VNA). As far as the indoor case is concerned, a shopping mall, an airport-in-check hall and the entrance hall of modern office building  have been considered. For the outdoor case, measurements have been conducted in suburban, residential and city center area's  environments. Path loss modeling is provided for directional and omnidirectional case as well as for the $2^{\text{nd}}$ and $3^{\text{rd}}$ strongest multipath components.
    \item
    An analysis of the channel angular spread is further provided. Our results are based on the mean and standard deviation of the angular spread computed from normalized power and azimuth angle of arrival of the MPCs.
    \item
    We finally provide results on the channel sparsity  by means of the the Gini index.
\end{itemize}

The remainder of this paper is structured as follows: Section~\ref{sec:ChannelMeasurements} presents the measurement setups and sets which were applied. Section~\ref{sec:Pathloss} presents the derived path loss models assuming indoor and outdoor scenarios as well as LoS and NLoS propagation. 
Sections~\ref{Sec:PAS} and~\ref{Sec:Sparsity} present results for the power angular spread and the channel sparsity, respectively,  
whereas Section~\ref{Sec:Conclusion} concludes the paper.

\section{Radio Channel Measurements}\label{sec:ChannelMeasurements}

The channel sounder employed in the radio channel measurements is discussed in~\cite{FrancisISWCS21}. There are some differences in the sounder parameters used in each scenario as listed in Table~\ref{tab:MeasurementDetails}. The descriptions of each measurement site are elaborated in the following subsections. 
\begin{figure*}[ht]
	\begin{center}
	\includegraphics[keepaspectratio,width=\linewidth]{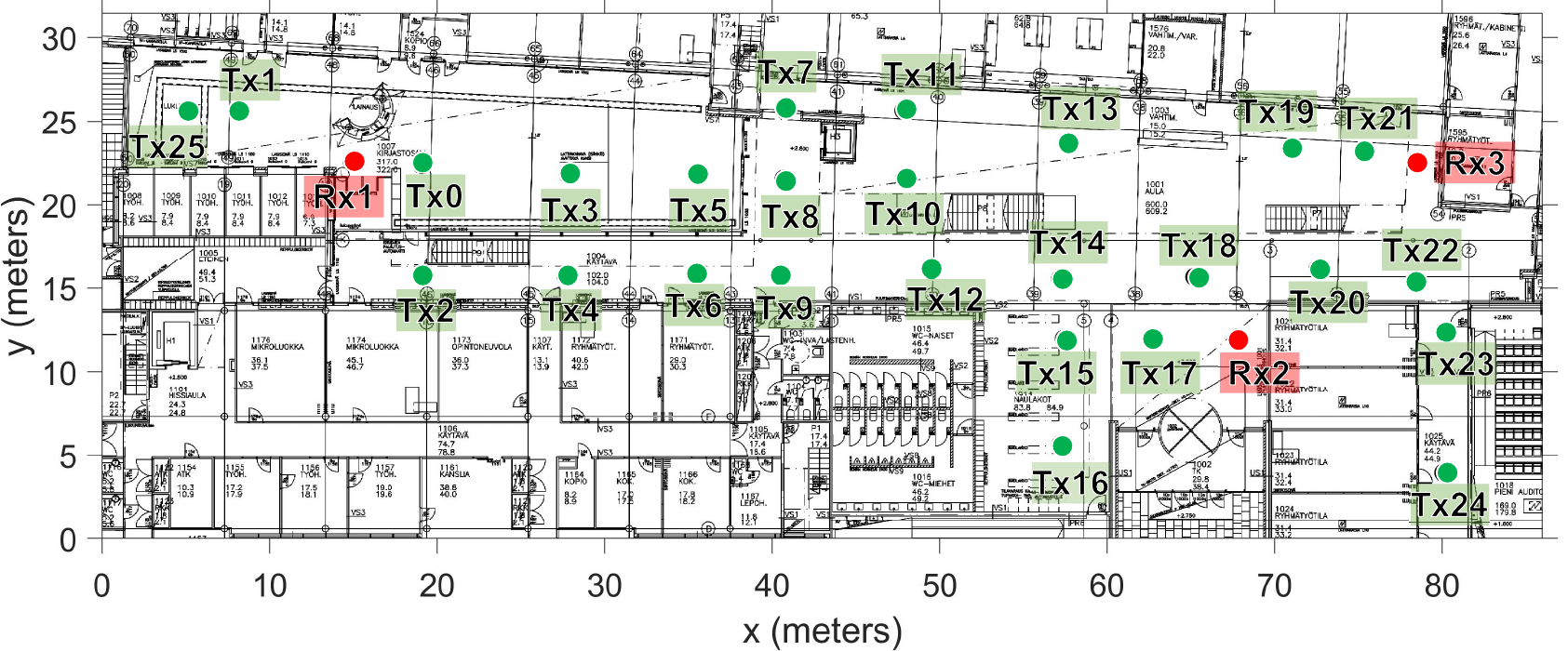}
	\caption{Antenna locations of the second measurement campaign in the entrance hall.}\label{fig:TUAS_map2}
	\end{center}
	\end{figure*}

\subsection{Indoor Measurement Campaign}\label{sec:MeasurementsIndoor}

The measurement campaigns performed in the shopping mall and airport check-in hall are described in \cite{Nguyen}. The third indoor site, which is further described in \cite{deGuzman22_VTCS}, is in the entrance hall of the Electrical Engineering building of Aalto University in Maarintie 8, Espoo, Finland. Two measurement campaigns were performed in this site and are referred in Table~\ref{tab:MeasurementDetails} as Entrance Hall 1 and Entrance Hall 2. The antenna locations of the two measurement campaigns are shown in Fig.~\ref{fig:TUAS_map1} and Fig.~\ref{fig:TUAS_map2}, respectively.


	

\subsection{Outdoor Measurement Campaign}
\begin{figure}[H]
\begin{center}
\includegraphics[keepaspectratio, width=\linewidth]{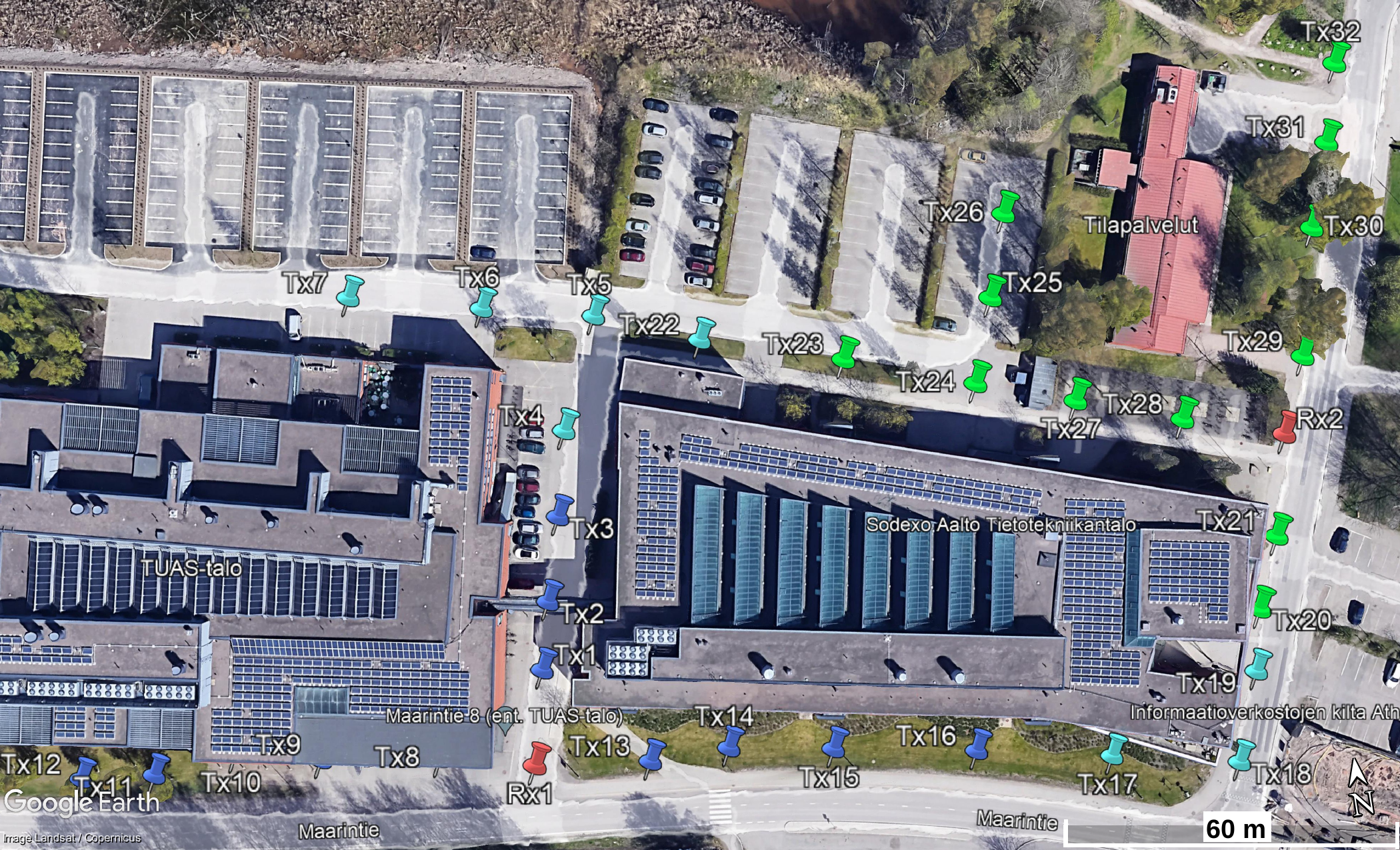}
\caption{Antenna locations of the measurement campaign in a suburban area. The blue, green, and cyan markers denote the Tx locations paired with Rx1, Rx2, and both Rx, respectively}\label{fig:campus_map}
\end{center}
\end{figure}

The first outdoor measurement campaign was conducted in a suburban area along Maarintie and Konemiehentie, Otaniemi, Finland. The outside walls of the buildings in the area are mainly made up of bricks and have glass windows and doors with metallic frames. There are also few large metallic structures installed on the walls. Metallic posts, trees, and parking space with cars moving from time to time can also be found in the area. The antenna locations for this measurement campaign are shown in Fig.~\ref{fig:campus_map}.

\begin{figure}[H]
\begin{center}
\includegraphics[scale=0.27]{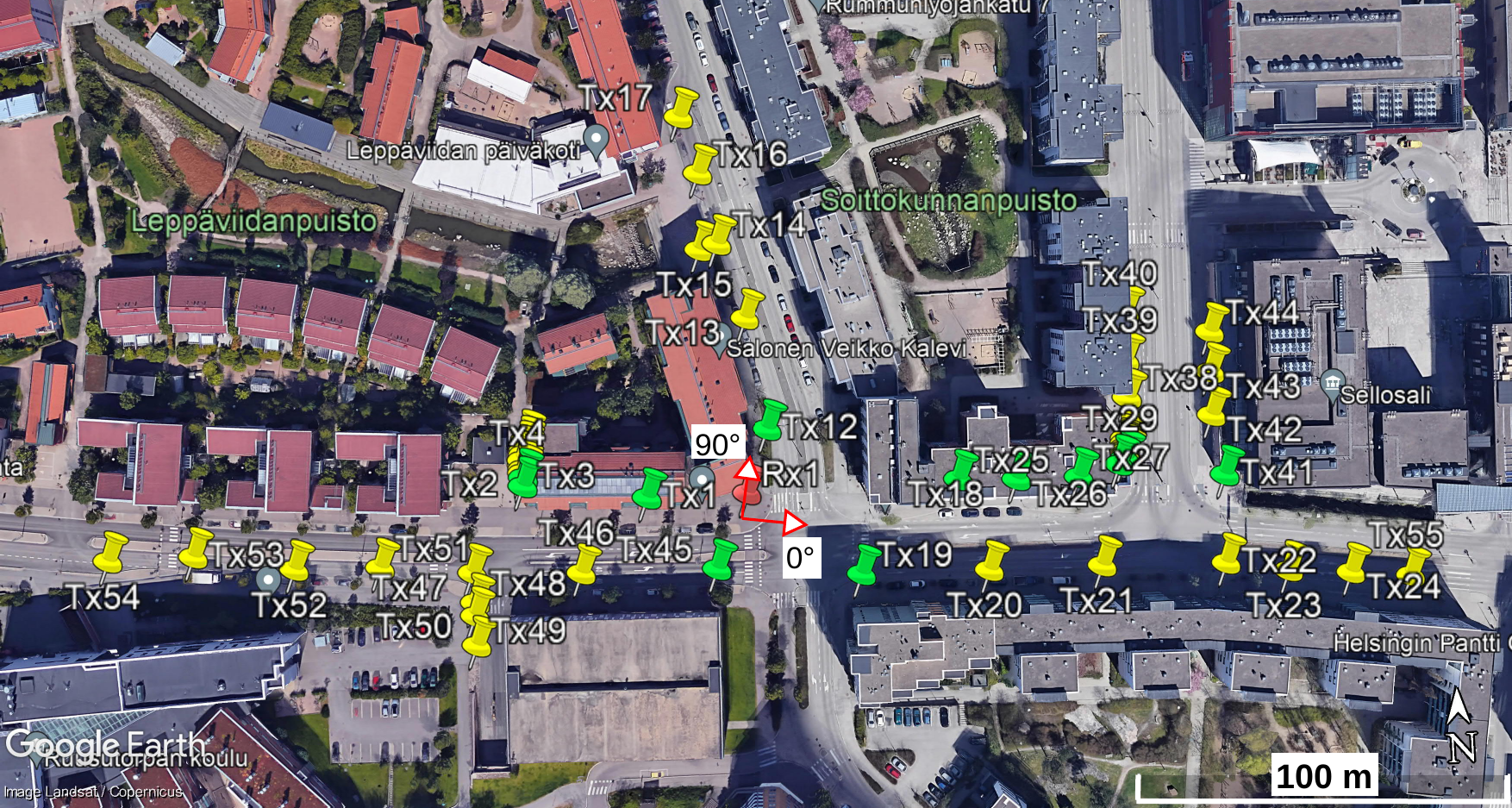}
\caption{Antenna locations of the measurement campaign in a residential area.}\label{fig:residential_map}
\end{center}
\end{figure}

The second measurement campaign was performed in a residential environment
along Leppavaarankatu, Espoo, Finland. The street is mostly surrounded by residential buildings and by some commercial buildings. Metallic street posts and trees can also be found in the area. The antenna locations 
for this residential area are shown in Fig.~\ref{fig:residential_map}.

\begin{figure}[H]
\begin{center}
\includegraphics[scale=0.37]{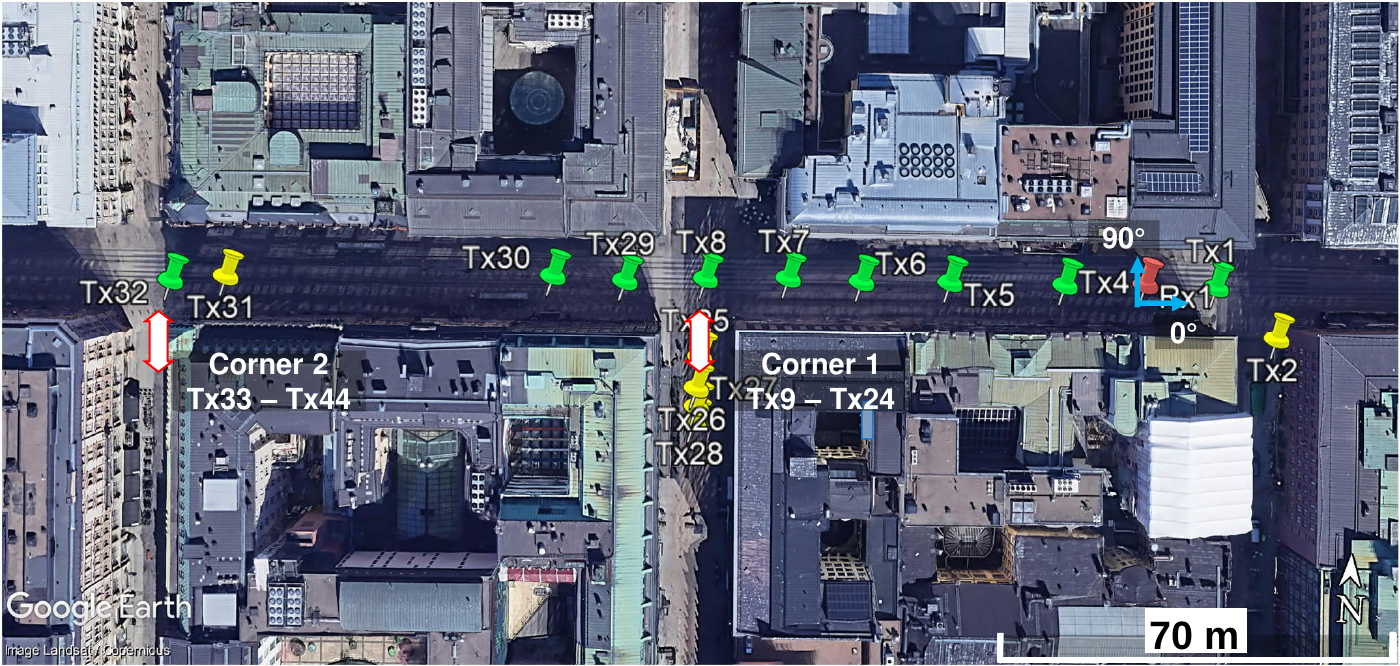}
\caption{Antenna locations of the measurement campaign in a city center.}\label{fig:city_center_map}
\end{center}
\end{figure}

The last outdoor measurement campaign was performed in an urban environment in the city center along Aleksanterinkatu, Helsinki, Finland. The street is surrounded by commercial buildings on both sides, forming a street canyon. The street is primarily intended for pedestrians, blocking vehicular traffic except for trams. There are rare metallic sign posts found on
the street. The location has heavy loads of pedestrians and has some vehicles that park on the side walks occasionally. The antenna locations 
for this residential area are shown in Fig.~\ref{fig:city_center_map}.

Closely-spaced antenna locations were positioned at the building corners $1$ and $2$. In all Tx-Rx links, particularly of outdoor scenarios, the presence of moving objects is inevitable during the measurement of directionally-resolved wideband channels. Moving objects affect only a portion of directionally-resolved channels and may block some propagation paths. Our measured channels are therefore snapshots of channels with moving objects. Still, the influence of path blockage due to moving objects is expected to be minor because of relative sparsity of propagation paths.

\begin{table*}[t]
	\begin{center}
		\caption{Measurement Campaign Details}\label{tab:MeasurementDetails}
		
		\begin{tabular}{|c|c|c|c|c|c|c|c|} \hline			
        \textbf{Measurement Detail} &\textbf{Shopping Mall} &\textbf{Airport} &\textbf{Entrance Hall 1} &\textbf{Entrance Hall 2} &\textbf{Suburban} &\textbf{Residential} &\textbf{City Center} \\ \hline
        RF (GHz) &$141.5$-$145.1$ &$141.5$-$145.1$ &$140$-$144$ &$140$-$144$ &$140$-$144$ &$140$-$144$ &$140$-$144$ \\ \hline
        Tx Antenna Height (m) &$1.89$ &\makecell{$1.7$ above \\ $2^{\rm nd}$ floor} &$1.85$ &$1.85$ &$1.85$ &$1.85$ &$2.00$  \\ \hline
        Rx Antenna Height (m) &$1.89$ &\makecell{$2.1$ above \\ $3^{\rm rd}$ floor} &$1.85$ &$1.85$ &$1.85$ &$1.85$ &$2.00$  \\ \hline
        EIRP (dBm) &$-12$ &$-12$ &$5$ &$5$ &$5$ &$5$ &$5$ \\ \hline
        Rx Azimuth Range ($^\circ$) &$0$-$360$ &\makecell{$0$-$25$, \\ $245$-$360$} &$40$-$250$ &\makecell{$-90$-$180$ (Rx1);\\$40$-$250$ (Rx2);\\$110$-$290$ (Rx3)} &$0$-$355$ &Mostly $0$-$355$ &Mostly $0$-$355$ \\ \hline
        Azimuth Step ($^\circ$) &$6$ &$5$ &$10$ &$5$ &$5$ &$5$ &$5$ \\ \hline
        Number of LOS Links &$16$ &$10$ &$2$ &$12$ &$32$ &$13$ &$19$ \\ \hline
        Number of NLOS Links &$2$ &$1$ &$9$ &$56$ &$8$ &$42$ &$21$ \\ \hline
        Environment Type &Indoor &Indoor &Indoor &Indoor &Outdoor &Outdoor &Outdoor \\ \hline
        Link Distance Range (m) &$3$-$65$ &$15$-$51$ &$3$-$47$ &$3$-$66$ &$2$-$172$ &$20$-$175$ &$10$-$178$ \\ \hline
		\end{tabular}
	\end{center}
\end{table*}

\section{Path Loss Modeling}\label{sec:Pathloss}
Hereafter, a brief description of the so-called close-in (CI) free space path loss model is presented. This model has been employed to fit experimental data to an empirical linear equation, based on the minimum mean square error (MMSE) estimation method.
 


The CI model is the most commonly used large scale path loss model\cite{RappaportCI}. This model has been extensively employed to predict signal strength  for a vast range of frequencies, such as \cite{Sun,Xing,Cheng}. 
According to this model, the RF signal strength as a function of the distance $d$ can be analytically expressed as \cite{C:XingRapaport}
\begin{align}\label{Eq:1}
{\text{PL}}^{\text{CI}}(f_c,d_{3d})= {\text {FSPL}}(f_c,d_0)+10n\log_{10}\left(\frac{d}{d_0}\right)
+\chi_\sigma^{\text{CI}}
\end{align}
where $d_0$ is the reference distance, $\text{FSPL} (f_c,d_0)$ is the large-scale free space path loss of the signal at carrier frequency $f_c$ over distance $d_0$, $n$ is the path loss exponent (PLE) and $\chi_\sigma^{\text{CI}}$ is the  shadow fading in dB which is modelled as a zero mean Gaussian random variable with a standard deviation of $\sigma$, expressed in dB\cite{J:SunShu}.

In what follows, path loss measurements and the corresponding modeling will be presented for both indoor and outdoor environments. Path loss modeling is performed taking into account: i) the strongest multipath component for every link, which will be referred to as directional modeling; ii) the sum of all multipath components for each Tx-Rx link, which will be referred to as omnidirectional modeling assuming non-coherent summation of multipath components; iii) the $2^{\text{nd}}$ and the $3^{\text{rd}}$ strongest multipath components for each link.

\subsection{Indoor path loss modeling}
Figs. 6 and 7 depict the results of path loss modelling assuming LoS and NLoS indoor links, respectively. Directional and omnidirectional path-loss modeling is provided as a function of distance $d$ by employing the CI free space path loss model. The same procedure has been repeated for the $2^{\text{nd}}$ and the $3^{\text{rd}}$ strongest MPCs of the received signal. Based on \cite[eqs. (30) and (32)]{J:SunShu}, path loss exponent, $n$, and standard deviation, $\sigma$, have been evaluated. Path-loss modeling parameters for the indoor measurements are summarized in Table ~\ref{Table:2}.

\begin{figure}[H]
	\includegraphics[width=\linewidth]{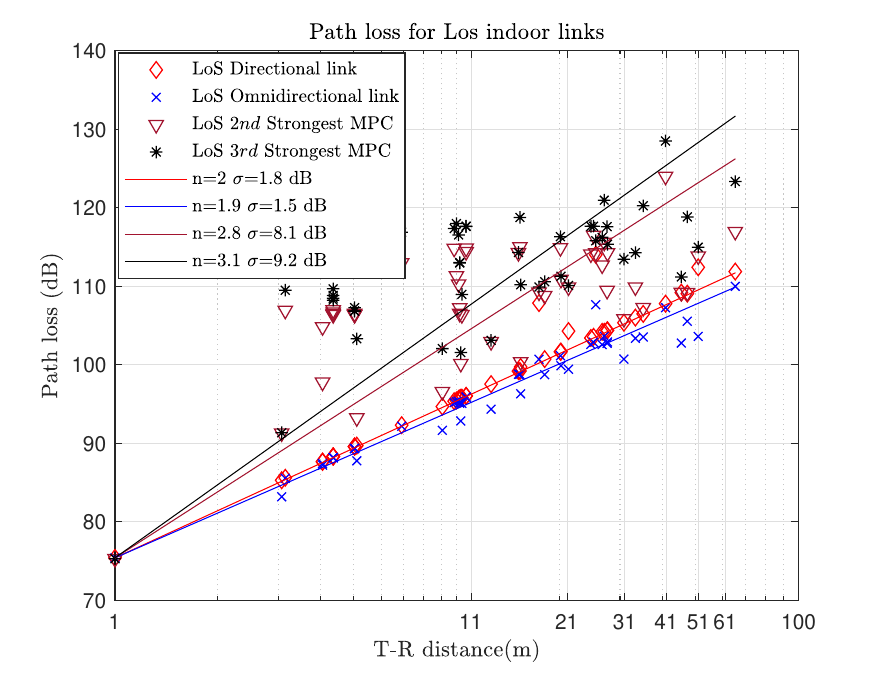}
	\centering
	\caption{Path loss modeling of LoS indoor links }
	\centering
	\label{Fig:14}
\end{figure}

\begin{figure}[H]
	\includegraphics[width=\linewidth]{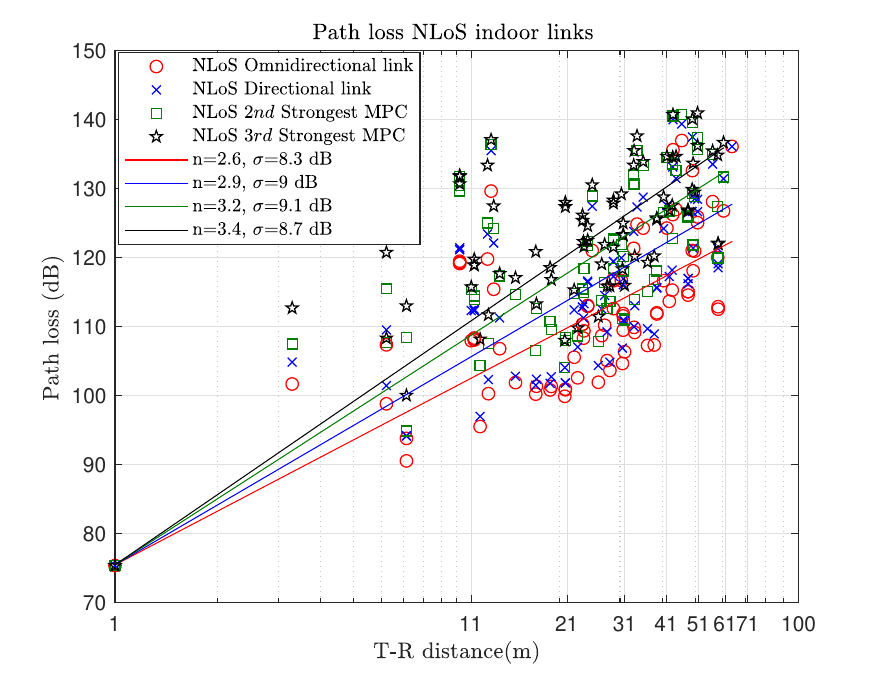}
	\centering
	\caption{Path loss modeling of NLoS indoor links }
	\centering
	\label{Fig:16}
\end{figure}

Specifically, as depicted in Fig. \ref{Fig:14} for LoS links directional modeling estimates PLE equal to $n =2 $ with standard deviation, $\sigma$, of 1.8 dB. Omnidirectional modeling, when calculated non-coherently as in our case, is expected to represent the best case in terms of signal losses which is verified by the estimated $n=1.9$ with standard deviation of $\sigma=1.5$ dB. On the other hand, the $2^{\text{nd}}$ and $3^{\text{rd}}$ strongest MPCs can be modeled by larger PLEs, i.e., $n=2.8$ and $n=3.1$, respectively. Moreover, they are characterized by quite large values of shadow fading, that is $\sigma=8.1$ dB for the $2^{\text{nd}}$ strongest MPC and $\sigma=9.2$ dB for the $3^{\text{rd}}$ strongest MPC.

\begin{small}
\begin{table}[H]
\caption{Indoor LoS and NLoS  path loss modeling}
  \begin{tabular}{|l|l|l|l|l|}
    \hline
    \multirow{1}{*}{Path loss modeling} &
      \multicolumn{2}{c}{LoS} &
      \multicolumn{2}{c|}{NLoS}\\
     & PLE (n) &  ($\sigma$)[dB] & PLE (n) & ($\sigma$)[dB]  \\
    \hline
    Directional & 2 & 1.8 & 2.9 & 9 \\
    \hline
    Omnidirectional & 1.9 & 1.5 & 2.6 & 8.3\\
    \hline
  $2^{\text{nd}}$ Strongest MPC & 2.8 & 8.1 & 3.2 & 9.1\\
  \hline
   $3^{\text{rd}}$ Strongest MPC & 3.1 & 9.2 & 3.4& 8.7\\
  \hline
  \end{tabular}
  \label{Table:2}
\end{table}
\end{small}

As expected, in Fig.\ref{Fig:16}, NLoS links are characterized by higher losses in all categories of path-loss modeling in Table ~\ref{Table:2} as expressed by the PLEs which are larger than the corresponding ones of the LoS case. In directional and omnidirectional modeling, significant increase is also observed in the shadow fading values which are now $\sigma=9$ dB and $\sigma=8.3$ dB, respectively. For the $2^{\text{nd}}$ and $3^{\text{rd}}$ strongest MPCs, shadow fading values are comparable to those of the LoS case.
Path-loss modeling parameters for the indoor measurements are summarized in Table ~\ref{Table:2}.

\subsection{Outdoor path loss modeling}
\begin{figure}[H]
	\includegraphics[width=\linewidth]{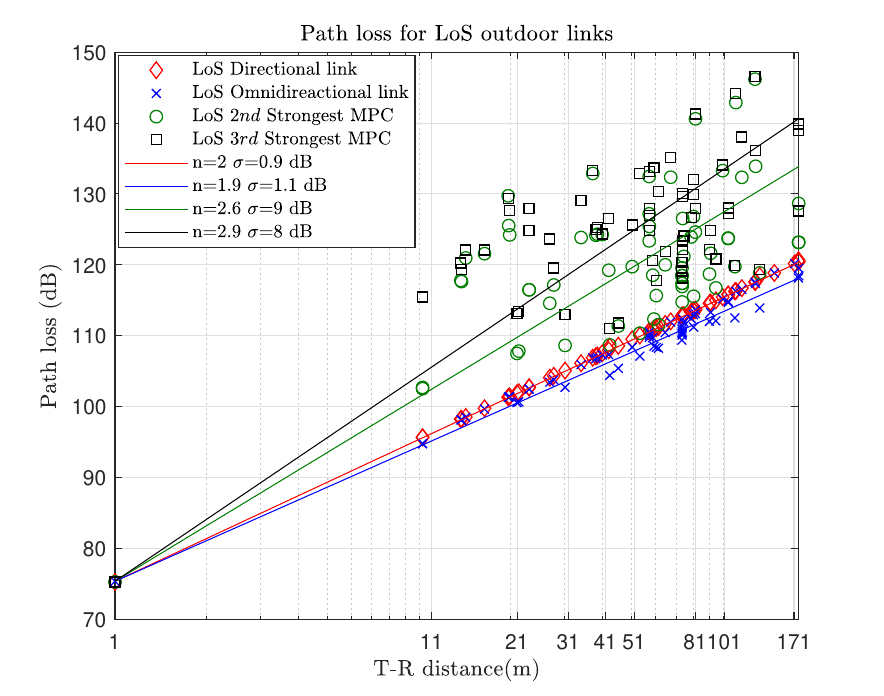}
	\centering
	\caption{Path loss modeling of outdoor LoS links }
	\centering
	\label{Fig:19}
\end{figure}

\begin{figure}[H]
	\includegraphics[width=\linewidth]{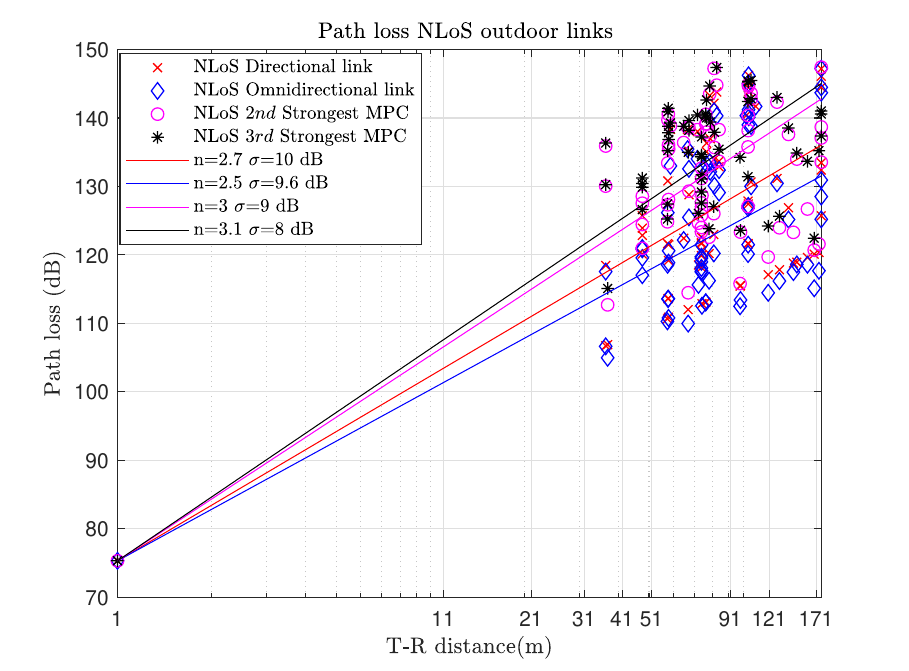}
	\centering
	\caption{Path loss modeling of outdoor NLoS links}
	\centering
	\label{Fig:15}
\end{figure}

Fig.~\ref{Fig:19} and Fig.~\ref{Fig:15} depict PLE as a function of distance $d$ after analyzing LoS and NLoS measurements of the outdoor scenarios. The same methodology was applied for outdoor measurements as in the indoor scenarios presented above. Thus, directional and omnidirectional path loss modeling is studied as well as for the $2^{\text nd}$ and $3^{\text rd}$ strongest MPCs. The corresponding results are available in Table ~\ref{Table:1}. 

For LoS propagation and directional modeling, PLE is equal to $n=2.0$ with standard deviation of $\sigma=0.9$ dB. Non-coherent omnidirectional modeling yields a lower PLE, namely $n=1.9$, with a slightly larger $\sigma$ of 1.1 dB compared to the directional case. The PLE values are quite similar to the corresponding ones for the indoor scenarios, while $\sigma$ takes lower values in the outdoor case.  The $2^{\text nd}$ and $3^{\text rd}$ strongest component analysis results in PLE of $n=2.6$ and $n=2.9$ with $\sigma=9$ dB and $\sigma=8$ dB, respectively. Compared to indoor LoS links, PLEs have slightly lower values, while standard deviations are similar.
 


Fig.~\ref{Fig:15} depicts the results for NLoS propagation. As expected, larger values of the PLE have been obtained (see Table~\ref{Table:1}), compared to LoS conditions, taking values from 2.6 up to 3.4 depending on modeled path loss types. Moreover, $\sigma$ is significantly increased taking values from 8 dB to 10 dB. In comparison with indoor results, outdoor NLoS propagation is characterized by slightly lower PLEs, while $\sigma$ values appear slightly larger.

\begin{small}
\begin{table}[ht]
\caption{Outdoor LoS and NLoS  results for CI path  loss model}
  \begin{tabular}{|l|l|l|l|l|}
    \hline
    \multirow{1}{*}{Path loss modeling} &
      \multicolumn{2}{c}{LoS} &
      \multicolumn{2}{c|}{NLoS}\\
     & PLE (n) &  ($\sigma$)[dB] & PLE (n) &  ($\sigma$)[dB]  \\
    \hline
    Directional & 2 & 0.9 & 2.7 & 10\\
    \hline
    Omnidirectional & 1.9 & 1.1 & 2.5 & 9.6\\
  \hline
  Directional $2^{\text{nd}}$ Strongest & 2.6 & 9 & 3 & 9\\
  \hline
  Directional $3^{\text{rd}}$ Strongest & 2.9 & 8 & 3.1 & 8\\
  \hline
  \end{tabular}
  \label{Table:1}
\end{table}
\end{small}


 

\section{Power Angular Spread Modeling}\label{Sec:PAS}
In this section, the estimation of angular power spread will first be briefly described, based on the mean and standard deviation of the angular spread computed from normalized power and azimuth angle of arrival of the MPCs.
Based on measured AoA and the corresponding power of the MPCs, the azimuth spread of arrival (ASA), $S_A$, can be calculated as the root-mean-square of angle spread, $\sigma$, of the strongest MPCs. 

The azimuth spread of the three strongest MPCs can be evaluated as \cite[eqs. (6.59) and  (6.60)]{Molischbook},\cite{Fleury, Gomez},
\begin{equation}
   S_{A}=\sqrt{\frac{\sum_{i=1}^{i=3}\left|e^{\imath\phi_i}-\mu_{AsA}\right|^2P_i}{\sum_{i=1}^{i=3}P_i}}
    \label{Eq:Sa}
\end{equation} 

\begin{equation}
    \mu_{ASA}=\frac{\sum_{i=1}^{i=3}e^{\left(\imath\phi_i\right)}P_i}{\sum_{i=1}^{i=3}P_i}
   \label{Eq:MuASA}
\end{equation}
where $\phi_i$ is the azimuth angle of arrival of the $i-th$ MPC considering values in the range $0$ to $2\pi$ and $P_i$ is the corresponding power gain and $\mu_{AsA}$ is the mean of ASA. Both equations are applied for every Tx-Rx antenna link.

Using 
 \cite[eqs.(2) and  (3)]{J:Zhang}, ASA values have been calculated for every Tx-Rx link of the available measurement sets. Then, the mean and standard deviation of $S_{A}$ have been calculated separately for LoS and NLoS links assuming indoor and outdoor environments, respectively. The corresponding results are given in Table~\ref{Table:Power Angular spread}. As it can be observed, a lowest mean value for $S_{A}$ of \ang{10} has been observed for NLoS links in outdoor environment, whereas a higher mean value of \ang{15.6} has been observed for NLoS links in indoor environment.

 \begin{center}
\begin{table}[H]
\caption{Power angular spread statistics in indoor and outdoor environments}
  \begin{tabular}{|l|l|l|l|l|}
    \hline
    \multirow{1}{*}{Environment} &
      \multicolumn{2}{c}{LoS} &
      \multicolumn{2}{c|}{NLoS}\\
     & mean ($S_{A}$) & std ($S_{A}$) & mean ($S_{A}$) & std ($S_{A}$)  \\
    \hline
    Indoor & \ang{13} & \ang{11.3} & \ang{15.6} & \ang{13.1}\\
    \hline
    Outdoor & \ang{10} & \ang{8.7} & \ang{10.1} & \ang{12.1}\\
  \hline
  
  \end{tabular}
  \label{Table:Power Angular spread}
\end{table}
\end{center}
 
 Note that the observed values are roughly in agreement with the statistics of directional channels available in \cite{Molischbook}, where mean $S_{A}$ values of less than \ang{20} have been reported for indoor and urban environments. It should be pointed out, however, that the results presented in \cite{Molischbook} are valid for below 6 GHz wireless channels.

\begin{figure}[H]
	\includegraphics[width=\linewidth]{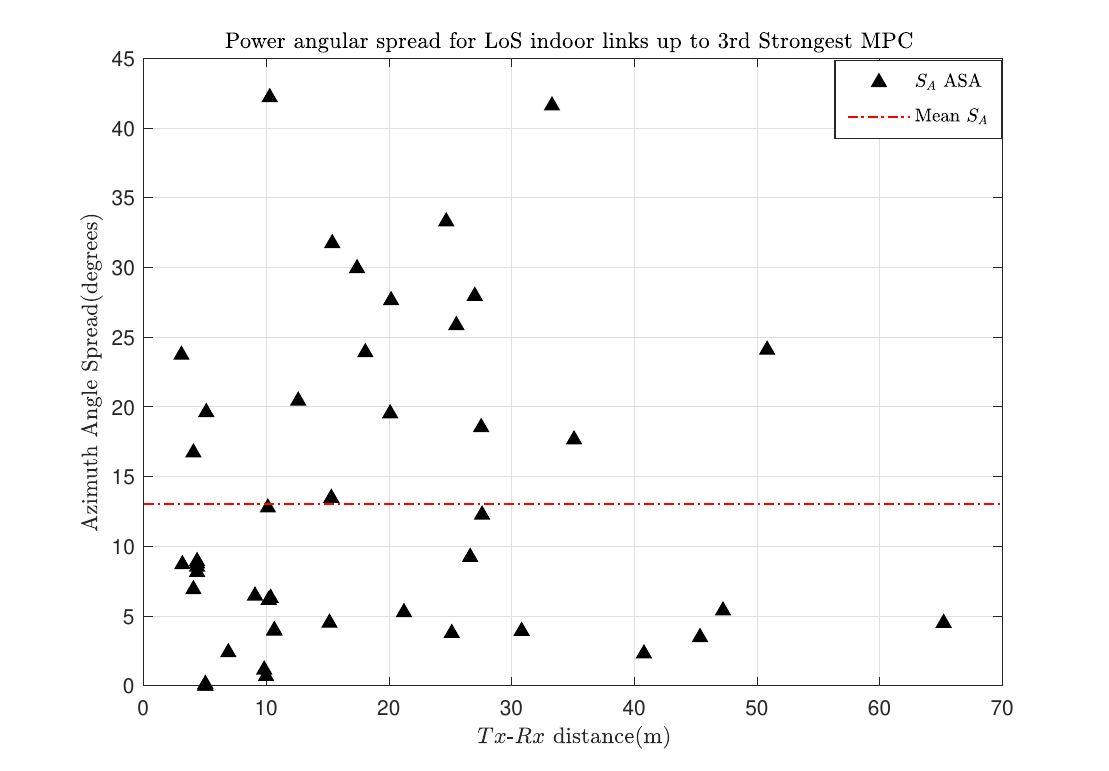}
	\centering
	\caption{Power angular spread for LoS indoor links }
	\centering
	\label{Fig:paindoorL}
\end{figure}

\begin{figure}[H]
	\includegraphics[width=\linewidth]{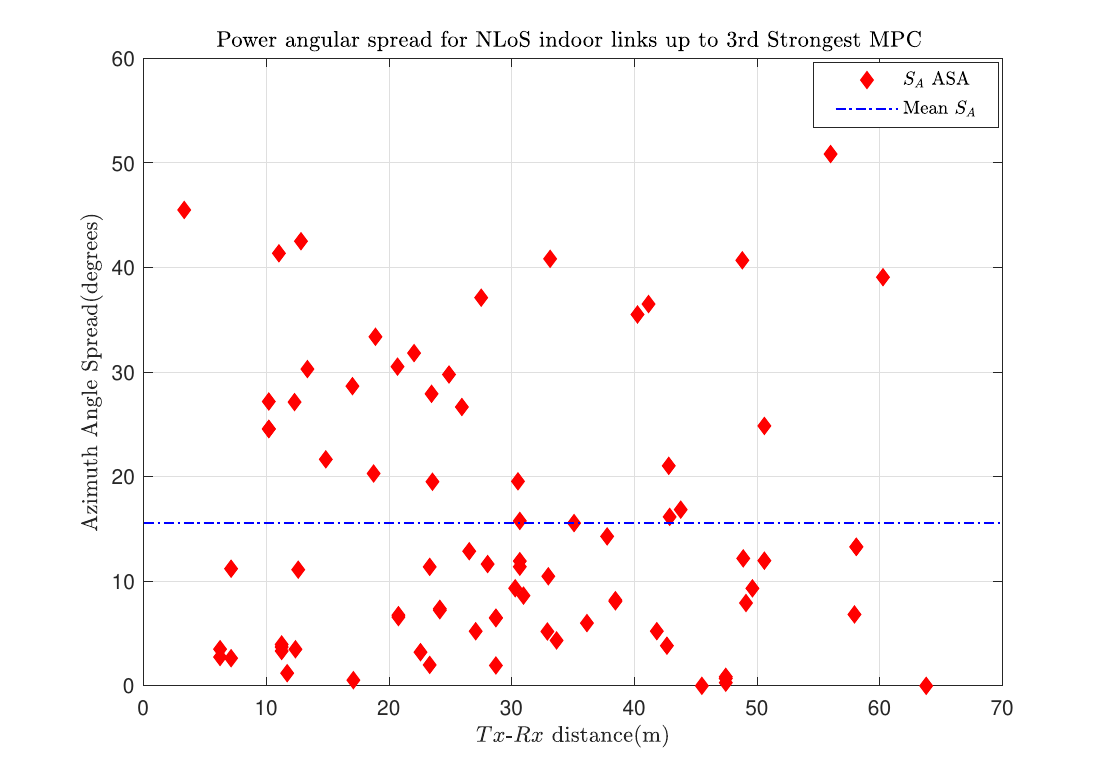}
	\centering
	\caption{Power angular spread for NLoS indoor links }
	\centering
	\label{Fig:paindoorNL}
\end{figure}
Figs.~\ref{Fig:paindoorL} and~\ref{Fig:paindoorNL} depicts the PAS for LoS and NLoS indoor links, respectively, based on measurements taken in the Shopping Mall, the Airport and Hall 1 and Hall 2 described in Table~\ref{tab:MeasurementDetails}. 
 As it can be observed, for the NLoS cases, azimuth angle spread $S_A$ has larger values i.e. up to approximately \ang{52}, as compared to the corresponding LoS cases.

\begin{figure}[H]
	\includegraphics[width=\linewidth]{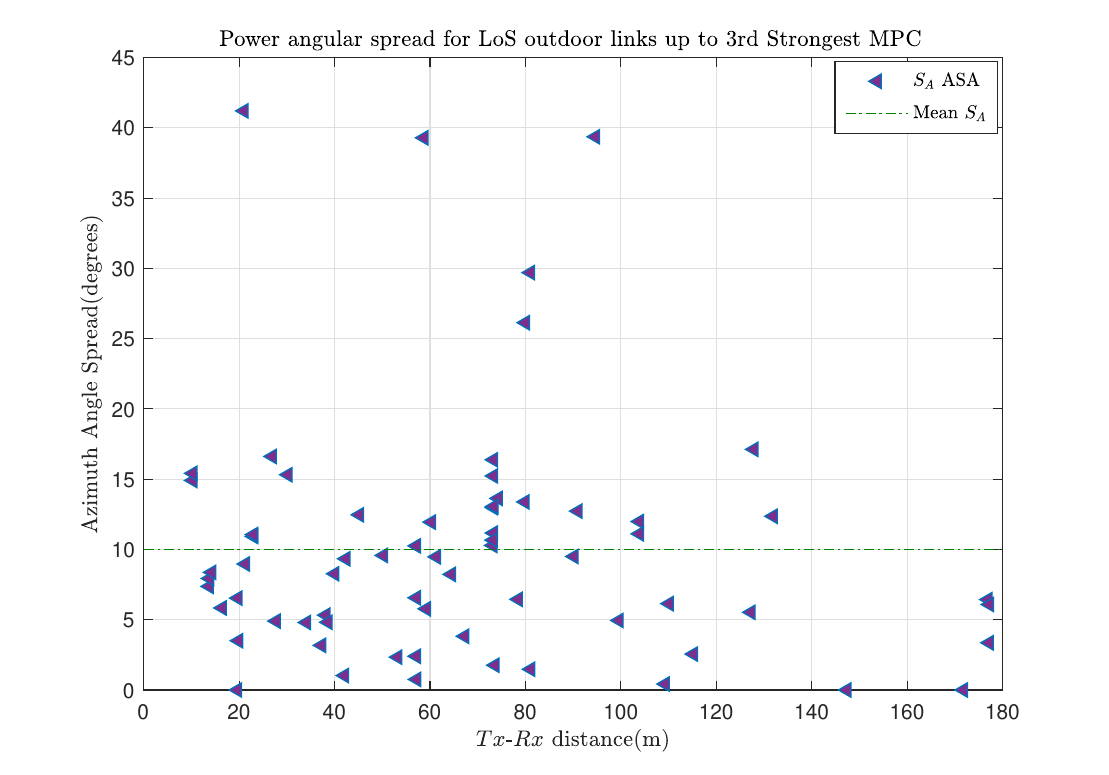}
	\centering
	\caption{Power angular spread for LoS outdoor links }
	\centering
	\label{Fig:paoutdoorL}
\end{figure}

\begin{figure}[H]
	\includegraphics[width=\linewidth]{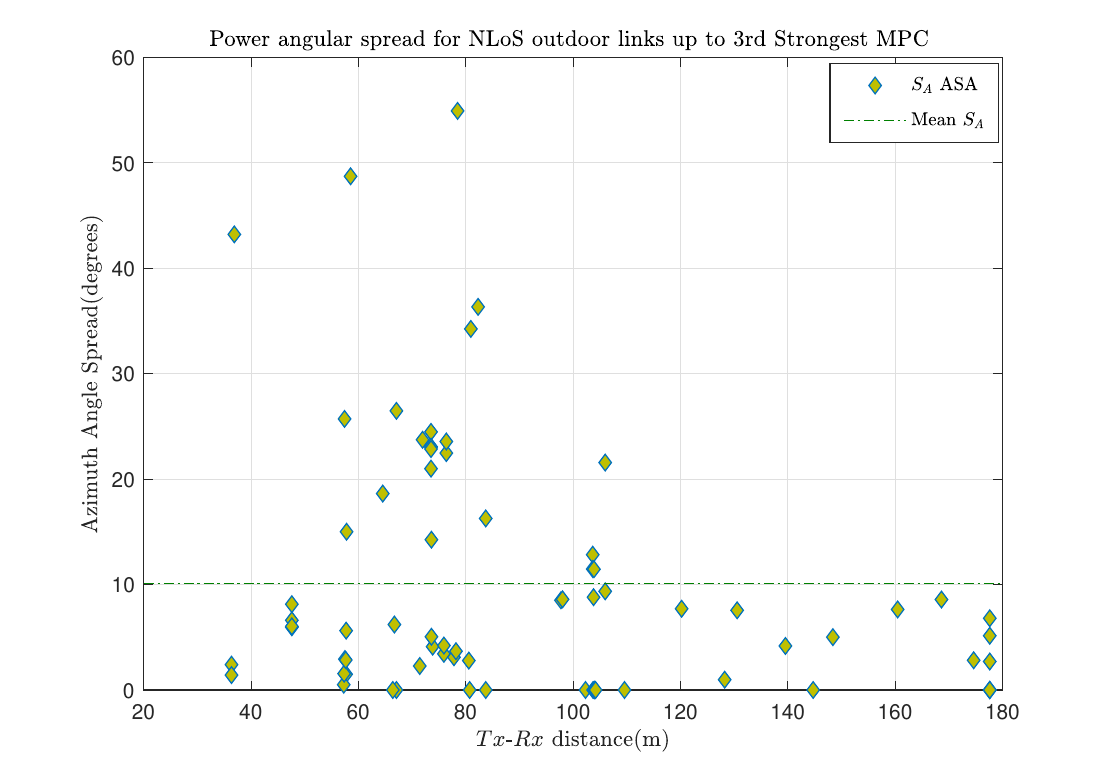}
	\centering
	\caption{Power angular spread for NLoS outdoor links }
	\centering
	\label{Fig:paoutdoorNL}
\end{figure}

Figs.~\ref{Fig:paoutdoorL} and~\ref{Fig:paoutdoorNL} depict  power angular spread values assuming outdoor LoS and NLoS links, respectively. As it can be observed, $S_A$ values for NLoS cases have greater values than the corresponding ones in LoS scenarios. Specifically, a maximum value of up to approximately \ang{55} has been observed in the former test case. Interestingly, our derived results indicate that 
there is no evidence that the angular spread depends on the distance, for all considered test cases.
This finding is also verified by the observation of Figs~\ref{Fig:paindoorL}-~\ref{Fig:paoutdoorNL}. 

Table~\ref{Table:Power Angular spread} summarizes the 
power angular spread statistics assuming both indoor and outdoor environments. 
As it can be observed, angular spread is larger in indoor environments as compared to outdoor ones. The corresponding  mean values of $S_{A}$ are \ang{13} vs \ang{10} for LoS links and \ang{15.6} vs \ang{10.1} for NLoS links. 
Additionally, it is evident that the mean value of $S_{A}$ takes larger values for NLoS links as compared to the LoS ones for indoor measurements. Specifically, the corresponding values are \ang{15.6} and \ang{13} for NLoS and LoS links, respectively. As far as the outdoor environments are concerned, it can be observed that the corresponding values are very close to each other, i.e., \ang{10.1} and \ang{10} for NLoS and LoS scenarios, respectively. Finally, it is evident that the standard deviation of $S_{A}$ is concentrated within a narrow interval from \ang{8.7} to \ang{13.1}, concluding that the corresponding values are very close for indoor and outdoor environments, as well as LoS and NLoS links. 

\section{Characterization of the Channel Sparsity}\label{Sec:Sparsity}
Hereafter, we first introduce the Gini index (GI) as a measure of the channel sparsity. 
Based on \cite{ZonoobiGini}, the  Gini index of a  path gain vector $P=[P_1,P_2,...,P_i]$, with its elements re-ordered and organized in ascending order i.e., $p_1<p_2<...,p_n$, is defined as
\begin{equation}
    \text{GI} \triangleq 1-2\sum_{i=1}^{N}\frac{|p_i|}{||P||_1}\left(\frac{N-i+\frac{1}{2}}{N}\right)
    \label{Giniequation}
\end{equation}
where $N$ is the number of MPCs of the wireless channel, $||\cdot||_1$ is the $L_1$ norm of the power vector and $i$ is the index of the sorted MPC.
Note that $\text{GI}$ is in the range between 0 and 1. A value of $\text{GI} = 0$ refers to the least sparse channel, e.g. when the energy is equally distributed to every MPC. On the other hand, a value of $\text{GI} = 1$ refers to a sparse channel in which the energy is being concentrated in only one component.

\begin{centering}
\begin{table}[ht]
\centering
\caption{Channel sparsity mean values in indoor and outdoor environments}
  \begin{tabular}{|l|l|l|l|l|}
    \hline
    \multirow{1}{*}{Environment} &
      \multicolumn{2}{c}{Number of MPCs per Antenna link}&
      \multicolumn{2}{c|}{Mean $GI$}\\
     & LoS & NLoS & LoS & NLoS  \\
    \hline
    Indoor & 21 & 28 & 0.93 & 0.92\\
    \hline
    Outdoor & 15 & 8 & 0.86 & 0.71\\
  \hline
 
 \end{tabular}
 \label{MPCsTable}
 
\end{table}

 \end{centering}

In what follows, using \eqref{Giniequation}, the sparsity of the considered wireless channel will be estimated by means of GI while the relevance with the number of MPCs will be also investigated. Additionally, it is expected that the existence of a strong LoS component, i.e. in the case of LoS links, should result in higher values of the GI (i.e. a more sparse channel) as compared to the case of NLoS links. 

\begin{figure}[H]
	\includegraphics
 [width=\linewidth]
 {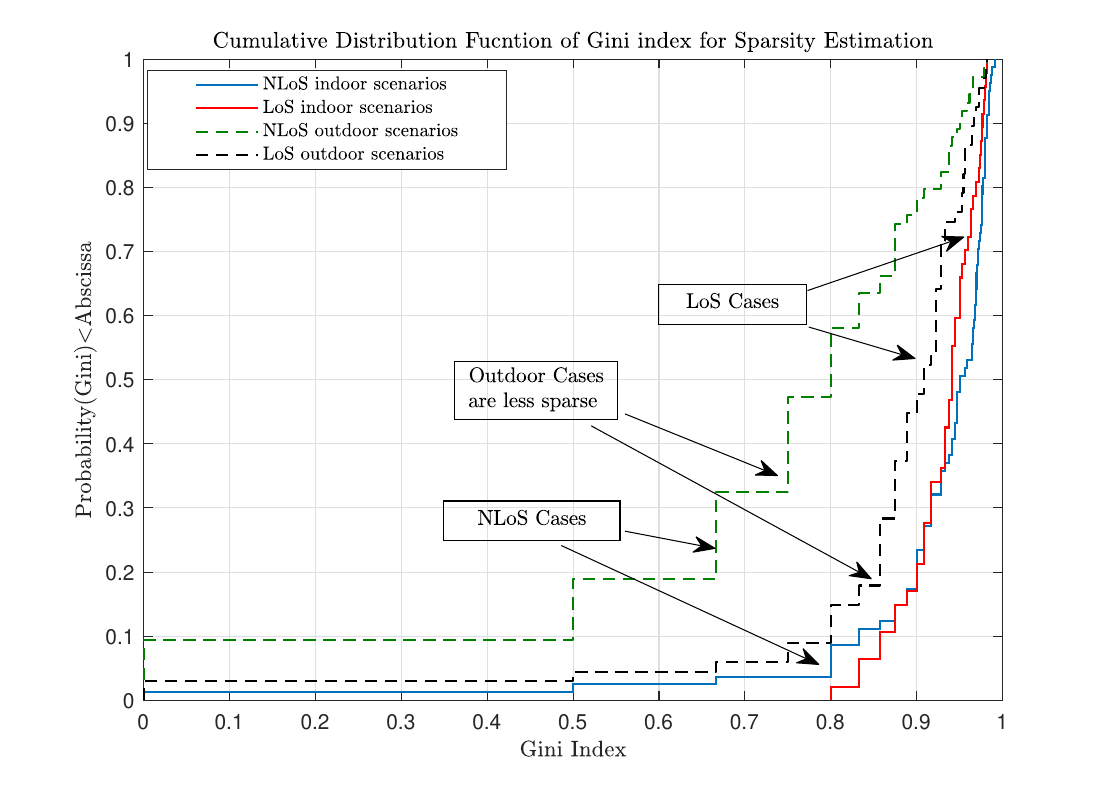}
	\centering
	\caption{Cumulative distribution function of Gini index }
	\centering
	\label{Fig:Gini}
\end{figure}

Table~\ref{MPCsTable} presents numerical results on the channel sparsity where the computed values of the GI are within the range 0.71 up to 0.93. Thus, we can state that the D-band wireless channels under investigation, are sparse channels. Moreover, it is evident that the number of MPCs do not affect the GI value for all considered test cases. This finding is also in agreement with the results reported in \cite{ZhangGini}. Finally, Fig.~\ref{Fig:Gini} depicts the cumulative distribution of the GI for all considered propagation scenarios and as it is evident, outdoor wireless channels are less sparse as compared to the indoor ones. 

Moreover, it can be observed that, as expected from Table~\ref{MPCsTable}, LoS channels are more sparse than the NLoS ones for every test case.

Finally, it is worth pointing out that by comparing Tables \ref{Table:2} and \ref{Table:1}, it can be observed that in the outdoor propagation scenarios, the values of path loss exponent, $n$, for the $2^{\text{nd}}$ and $3^{\text{rd}}$ strongest MPC are lower than those of the indoor scenarios. 
 In other words, the attenuation of the path loss for the $2^{\text{nd}}$ and $3^{\text{rd}}$ strongest components for outdoor cases is less severe  
indicating a less sparse channel compared to indoor cases. 
This conclusion is also verified by the results presented in Table~\ref{MPCsTable} and Fig.\ref{Fig:Gini}.


\section{Conclusion}\label{Sec:Conclusion}

In this paper, we first presented new results on path loss modeling at 142 GHz, based on measurement campaigns conducted in both indoor and outdoor environments, thus providing estimations for path loss exponent and shadow fading. Path loss modeling has been also applied for the $2^{\text{nd}}$ and $3^{\text{rd}}$ strongest MPCs of the received signal, in order to obtain further insights on the spatial characteristics of the channel. The obtained values of the path loss exponent are in the range of 2 to 3.1 and 2 to 2.9 for indoor and outdoor scenarios, respectively. 
Note that the estimated values for the path loss exponent  are pretty close to those reported for currently used RF-frequencies, thus indicating that sub-THz links may experience a similar propagation behaviour. For all considered propagation scenarios, results on the power angular spread have also been presented, by taking into account the 3 strongest MPCs. The corresponding calculated mean values lie within \ang{10} and \ang{15.6} and thus, the wireless channel can be characterized as directional. Moreover, there is no strong indication that the distance between Tx-Rx affects the power angular spread. 
Finally, results on the sparsity of the wireless channel have been reported using the Gini index, thus providing further insights as to the factors affecting wireless propagation on the considered frequencies.

\section*{Acknowledgement}
The presented work has been performed within the frame of ARIADNE project which is a three years Research and Innovation action / project under the EU programme Horizon 2020 (Grant Agreement no. 871464).
The publication of the article in OA mode was financially supported by HEAL-Link.

\bibliographystyle{IEEEtran}
	\bibliography{IEEEabrv,bilbiography2}

\end{document}